# Optical excitation of electromagnons in hexaferrite

Hiroki Ueda[1,*], Hoyoung Jang[2], Sae Hwan Chun[2], Hyeong-Do Kim[2], Minseok Kim[2], Sang-Youn Park[2], Simone Finizio[1], Nazaret Ortiz Hernandez[1], Vladimir Ovuka[3], Matteo Savoini[3], Tsuyoshi Kimura[4], Yoshikazu Tanaka[5], Andrin Doll[1], and Urs Staub[1,*]

[1] *Swiss Light Source, Paul Scherrer Institute, 5232 Villigen-PSI, Switzerland.*

[2] *PAL-XFEL, Pohang Accelerator Laboratory, Pohang, Gyeongbuk 37673, South Korea.*

[3] *Institute for Quantum Electronics, Physics Department, ETH Zurich, 8093 Zurich, Switzerland.*

[4] *Department of Advanced Materials Science, University of Tokyo, Kashiwa, Chiba 277-8561, Japan.*

[5] *RIKEN SPring-8 Center, Sayo, Hyogo 679-5148, Japan.*

**Abstract**: Understanding ultrafast magnetization dynamics on the microscopic level is of strong current interest due to the potential for applications in information storage. In recent years, the spin-lattice coupling has been recognized to be essential for ultrafast magnetization dynamics. Magnetoelectric multiferroics of type II possess intrinsic correlations among magnetic sublattices and electric polarization (*P*) through spin-lattice coupling, enabling fundamentally coupled dynamics between spins and lattice. Here we report on ultrafast magnetization dynamics in a room-temperature multiferroic hexaferrite possessing ferrimagnetic and antiferromagnetic sublattices, revealed by time-resolved resonant x-ray diffraction. A femtosecond above-bandgap excitation triggers a coherent magnon in which the two magnetic sublattices entangle and give rise to a transient modulation of *P*. A novel microscopic mechanism for triggering the coherent magnon in this ferrimagnetic insulator based on the spin-lattice coupling is proposed. Our finding opens up a novel but general pathway for ultrafast control of magnetism.

[*] To whom correspondence should be addressed: hiroki.ueda@psi.ch and urs.staub@psi.ch

Magnetoelectric multiferroics of type II commonly exhibit magnetic order that induces ferroelectricity, and hence, there is naturally a strong coupling between magnetism and lattice, i.e., spin-lattice coupling. This class of materials has attracted enormous interest in the last two decades [1]. Interest in these multiferroics lies, for example, in electric (magnetic) manipulation of magnetism (polarization) [2-4] and in coupled spin-lattice excitations called electromagnons [5,6]. This inherent correlation between magnetism and ferroelectricity enables simultaneous control of principal excitation characters, magnons and phonons, in a wide dynamic range. However, time-resolved studies on the coupling, e.g., optical excitation of electromagnons and attempts to understand the microscopic origin of their creation, have been limited so far [7-10].

Optical manipulation of magnetic moments and identification of the underlying excitation mechanism are important objectives towards new-generation storage technologies and have been intensely debated in the context of spin-lattice coupling. These are, e.g., ultrafast non-thermal magnetization switching/reorientation via a change in magnetocrystalline anisotropy [11-13], a coherent excitation of magnons by driving an electromagnon or optical phonon mode [8,14,15], and an ultrafast demagnetization that generates a transverse strain wave via the Einstein-de Haas effect [16]. The importance of spin-lattice coupling is well established in multiferroics with their non-trivial magnetic orders resulting in remarkable magnetoelectric effects. Often, multiferroicity emerges due to magnetic frustration resulting in a high sensitivity of the property to bond angles. Tuning a lattice parameter, e.g., using strain, is indeed a standard approach to obtain multiferroicity [17]. The well-documented knowledge of spin-lattice coupling in multiferroics inspires us to explore a new pathway to create magnons via ultrafast modulation in the lattice other than known mechanisms, e.g., coherent displacive excitation [11] and the inverse Cotton-Mouton effect [18].

In this paper, we study the ultrafast magnetization dynamics in a room-temperature multiferroic Z-type hexaferrite $Sr_3Co_2Fe_{24}O_{41}$ with a ferrimagnetic (FM) and an antiferromagnetic (AFM) cycloidal sublattice [19,20]. The dynamics of the individual sublattices upon a femtosecond above-bandgap excitation were investigated by time-resolved resonant x-ray diffraction (tr-RXD). A coherent entangled magnon mode of the AFM and FM sublattices is observed and is described by the Landau-Lifshitz-Gilbert (LLG) equation as an electromagnon mode. Nonreciprocal directional dichroism (NDD) in microwave transmission spectra further confirms its origin as an electromagnon mode. An alternative excitation mechanism for the magnon creation based on the ultrafast Einstein-de Haas effect is proposed. Such a novel excitation mechanism for optical driving of magnetic modes possibly needs to be more generally considered in frustrated magnetic insulators.

Z-type hexaferrite $Sr_3Co_2Fe_{24}O_{41}$ crystals are commonly described in space group $P6_3/mmc$ with lattice constants of $a \approx 5.87$ Å and $c \approx 52.07$ Å [19]. The crystal structure, displayed in Fig. 1a, comprises two basic magnetic blocks, S blocks and L blocks, which alternately stack along [001]. Because of many magnetic Fe sites in the chemical unit cell, the magnetic structure of hexaferrites is described by a collinear-FM structure in each magnetic block. Arrows in Figs. 1b and 1c represent the local structures: a red (blue) arrow represents the magnetic moment of an L (S) block $\mu_L$ ($\mu_S$). There is magnetic frustration at the interface between S and L blocks, which is tunable by chemical substitution that changes the specific Fe-O-Fe bond angle (see the green dotted box in Fig. 1a) [21]. As a result, a non-collinear magnetic structure shown in Fig. 1b is stabilized below the multiferroic transition temperature $T_{MF} \approx 410$ K [20,22], whereas a collinear-FM structure shown in Fig. 1c appears above $T_{MF}$.

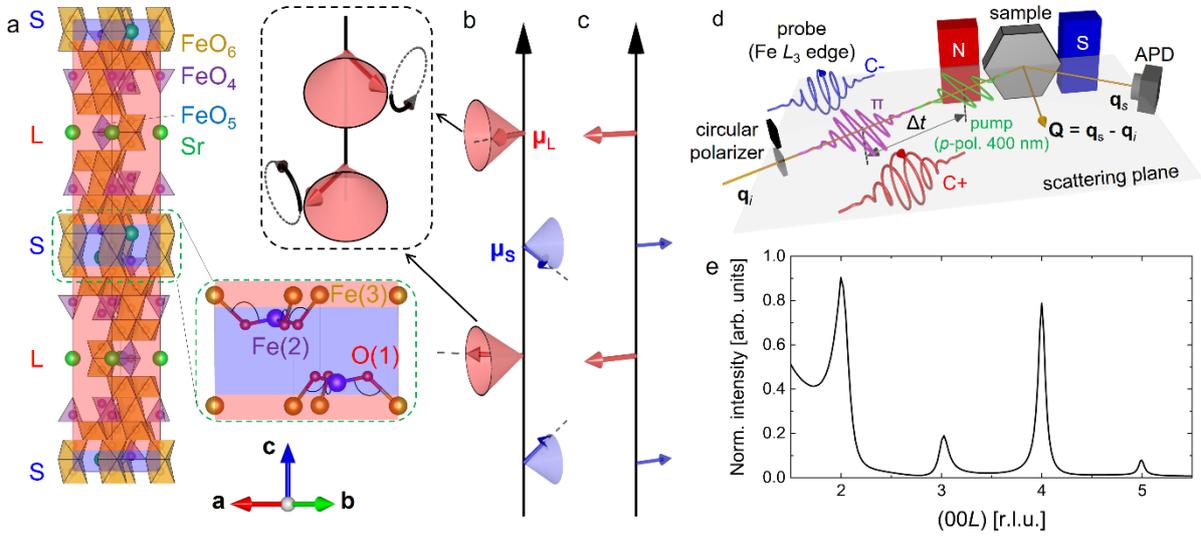

**Figure 1. Crystal structure and magnetic structures of Z-type hexaferrite $Sr_3Co_2Fe_{24}O_{41}$ and time-resolved resonant x-ray diffraction setup.** (**a**) Crystal structure of the Z-type hexaferrite $Sr_3Co_2Fe_{24}O_{41}$ drawn by VESTA [38]. $Co^{2+}$ sits on some $Fe^{3+}$ sites. A green dotted box shows an enlarged view of the local structure at the interface between adjacent magnetic blocks, where some atoms not relevant to the magnetic frustration are omitted. (**b,c**) Simplified magnetic structures of $Sr_3Co_2Fe_{24}O_{41}$ in (**b**) the conical phase and (**c**) the collinear-ferrimagnetic phase. Red and blue arrows denote net moments in magnetic L and S blocks ($\mu_L$ and $\mu_S$), respectively. In the black dotted box, the precession of $\mu_L$ in the magnetic excitation mode is shown by black arrows. Here $\mu_S$ is omitted because of its small precession amplitude. (**d**) Sketch of the experimental setup used for the time-resolved resonant diffraction experiments (details are described in Method). (**e**) Resonant diffraction profile from $Sr_3Co_2Fe_{24}O_{41}$ along (00$L$) in the conical phase.

The non-collinear transverse-conical structure is composed of a collinear-FM component with a propagation vector $\mathbf{k} = (0, 0, 0)$ and a non-collinear cycloidal component with $\mathbf{k} = (0, 0, 1)$. Hereafter, we refer to the former component as the FM component and to the latter as the AFM component. Whereas the FM component hosts magnetization ($M$), the AFM component gives rise to electric polarization (*P*) through spin-orbit couplings [19,23,24]. A previous RXD study on $Sr_3Co_2Fe_{24}O_{41}$ investigated the magnetic sublattices separately using two reflections sensitive to the individual sublattices [24] [see Fig 1e for a typical resonant diffraction profile along (00*L*) in the conical phase]. It revealed a strong coupling between the two sublattices in the adiabatic limit, essential for the magnetoelectric effect.

THz time-domain spectroscopy (THz-TDS) revealed the presence of an electromagnon mode around 1 THz [25,26], indicative of the magnetoelectric coupling on ultrafast timescales. This mode is the higher-frequency mode of the two possible electromagnon modes with AFM resonance obtained by solving the LLG equation. The lower-frequency mode though has not been reachable in the THz-TDS experiment. The expected dynamics of the slower mode are visualized in the dotted box of Fig. 1b, which involves anti-phase oscillation between adjacent two **μ**$_L$ and also between two **μ**$_S$ with two-order smaller motions for the latter. In addition, an even slower magnetic mode is observed by the microwave spectroscopy technique at GHz frequencies [27,28], which might be referred to as the toroidal-magnon mode or Nambu-Goldstone mode [29], the lowest-energy magnon mode in the transverse conical state.

**RESULTS**

We collected intensities of the (003) space-group forbidden and of the (004) allowed Bragg reflections sensitive to the AFM and the FM component, respectively, in addition to the fluorescence signal. The optical pump beam wavelength was 400 nm, which results in an above-bandgap excitation. Figure 2 shows tr-RXD intensities of the (003) and (004) Bragg reflections measured with π polarization at 709.6 eV, normalized by unperturbed intensities ($I_{OFF}$). Following the laser excitation, both reflections show (O-1) a rapid decrease within ~200 fs (see Figs. 2a and 2b), (O-2) oscillations (see Figs. 2c and 2d), (O-3) a slow response that is either a decrease [for (003) and (004) with low fluence] or an increase [for (004) with high fluence] (see Figs. 2e and 2f), and (O-4) a recovery following the precedent transient changes. The time traces of the normalized intensities ($I_{ON}/I_{OFF}$) are fitted by these four components defined as

$$\frac{I_{ON}}{I_{OFF}}(t) = -A_{rap}\left[1 - e^{-(t-t_0)/\tau_{rap}}\right] + A_{osc}e^{-(t-t_0)/\tau_{osc}}\cos[2\pi f t - \varphi] - A_{slow}\left[1 - e^{-(t-t_0)/\tau_{slow}}\right] + A_{rec}\left[1 - e^{-(t-t_0)/\tau_{rec}}\right] + 1. \quad (1)$$

Here $A_{rap}$ ($\tau_{rap}$), $A_{osc}$ ($\tau_{osc}$), $A_{slow}$ ($\tau_{slow}$), and $A_{rec}$ ($\tau_{rec}$) are amplitudes (relaxation times) of (O-1) the rapid decrease, (O-2) oscillation, (O-3) slow response, and (O-4) recovery, respectively, *f* is the oscillation frequency of the mode, $t_0$ is time zero, and $\varphi$ is a phase shift. The fit results in an oscillation frequency of $f_1 \approx 42.0 \pm 1.0$ GHz at room temperature (shown as translucent blue curves in Fig. 2c,d). Best fits for the parameters are listed in Supplementary Information.

The oscillations exhibit the same frequency for the two reflections, but they are opposite in phase. Because the (003) and (004) reflections sample the AFM and FM components, respectively, the oscillations with the same frequency indicate a coherent excitation of a magnon mode entangling both components. Cooling to 84 K shows qualitatively similar results except for; (I) $f_1$ shifts to a slightly higher frequency [~46.3±0.3 GHz (see the black translucent curves in Figs. 2c,d, representing the fitted oscillation component)], (II) the oscillation lives more protracted, and (III) the slow response of (004) turns from an increasing to decreasing behavior despite the equal fluence used (compare black and dark green curves).

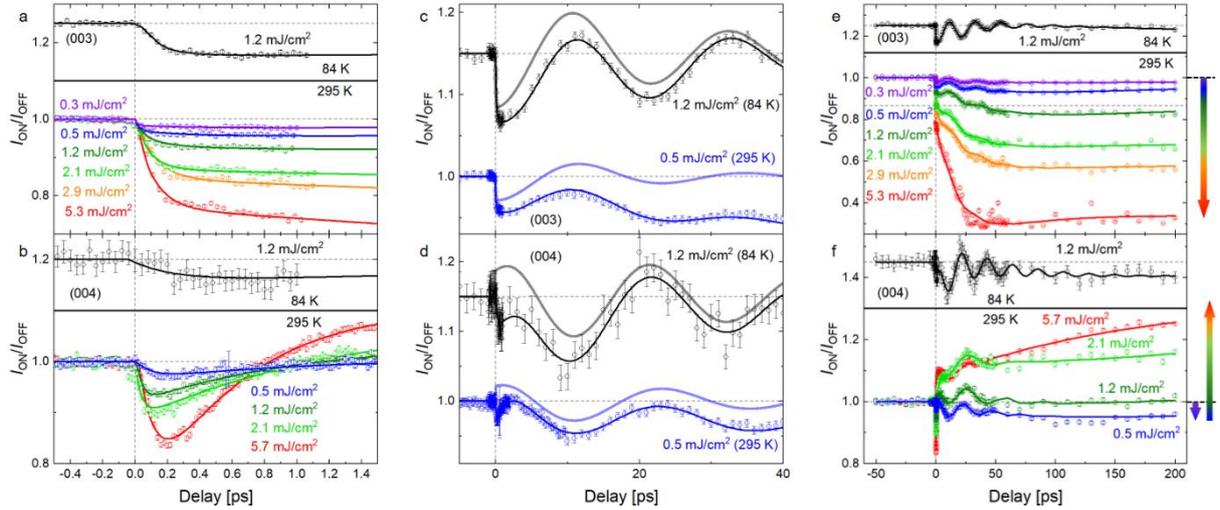

**Figure 2**. **Time traces of (003) AFM and (004) FM reflections.** (**a,c,e**) (003) AFM and (**b,d,f**) (004) FM reflections measured with π polarization of incident x-ray beams at 709.6 eV; (**a,b**) short time trace around $t_0$ (from −0.5 ps to 1.5 ps), (**c,d**) medium time range (−5 ps to 40 ps) to improve visibility of the intensity oscillations, and (**e,f**) extended time trace (from −50 ps to 200 ps). Each color corresponds to a different laser fluence or temperature depicted with the label of the same color, and solid curves are best fits. A translucent curve in (**c**) and (**d**) shows the oscillation component in the best fit. All data were taken at room temperature except for black curves taken at 84 K. Vertical and horizontal dotted lines indicate $t_0$ and the intensities before $t_0$, respectively.

Magnetic circular dichroic signals of the (004) reflection can directly capture $M$ [24] and its time evolution upon laser excitation. Figure 3 shows tr-RXD intensities taken with circular polarization at room temperature and extracted circular dichroic signals. An additional feature appears in the circular dichroic signals, (O-5) a significantly slower oscillation at $f_2 \approx 5.7±0.1$ GHz (see black data in Fig. 3a). The observation of a rapid decrease within < 200 fs followed by an increase of the overall signals indicates an ultrafast demagnetization followed by a delayed and slower increase of the FM component. The increase in the FM component goes hand in hand with the reduction of the AFM component (see Fig. 2e), suggesting a transformation of AFM to FM moment components.

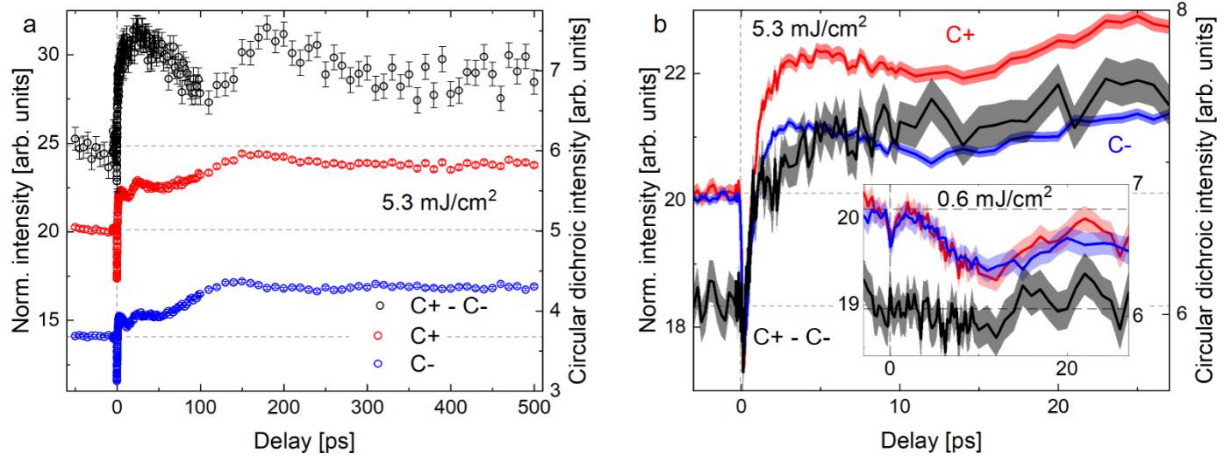

**Figure 3**. **Time traces of the circular dichroic (004) FM reflection.** (**a**) Extended time trace with high fluence (−50 ps to 500 ps, 5.3 mJ/cm$^2$) and (**b**) short time range around $t_0$ (−2 ps to 27 ps, 5.3 mJ/cm$^2$ and 0.6 mJ/cm$^2$), measured at room temperature. Red (blue) plots are normalized intensities taken with C+ (C−), while black plots are the difference between them, i.e., $I_{C+} - I_{C-}$. In (**b**), data taken with C− are vertically shifted by 6 [arb. units]. Inset shows data taken with low fluence (0.6 mJ/cm$^2$) for better visibility of the oscillation with $f_1$. Errors are shown as translucent filling in (**b**).

To better capture the modes observed in the time traces, we performed microwave spectroscopy measurements. Figure 4 shows the NDD signals of microwave transmission spectra in the relevant frequency ranges taken at room temperature and in various magnetic fields up to ±0.35 T. Resonant features with clear NDD are noticeable in both frequency ranges, consistent with $f_1$ and $f_2$. Furthermore, the NDD signals indicate both modes are electric- and magnetic-dipole active [6], meaning that these modes have an electromagnon character. Note that the higher-frequency $f_1$ mode exhibits extraordinary large NDD of up to 3.5 dB, which translates to a change in transmitted power > 100% and is much larger than the maximum change of ~11% for the lower-frequency $f_2$ mode. This implies a considerable resonator strength coupled to the electric field of the microwave and corresponding to significant modulations in $M$ and $P$ when the mode is resonantly excited. While NDD is less pronounced in the low-frequency range, microwave absorption is resonantly enhanced (see Supplementary Information), reminiscent of previous NDD/microwave studies on a chiral magnet [30]. All modes harden for increasing magnetic fields; resonance frequencies in the lower-frequency regime are almost linear in the amplitude of the static magnetic field consistent with the Zeeman effect, whereas that of the higher-frequency mode is strongly non-linear. Furthermore, the modes soften for elevated temperatures, as shown in Fig. S6 (see Supplementary Information). These results indicate that the oscillation with $f_2$ could be assigned to the toroidal magnon or Nambu-Goldstone mode, the lowest-energy magnon mode showing NDD [31], while the oscillation with $f_1$ would be assigned to a more complicated electromagnon excitation. Together with no evident heat-driven lattice expansion shown in SI, we conclude that the oscillations in tr-RXD are not caused by longitudinal strain waves but are of magnetic origin.

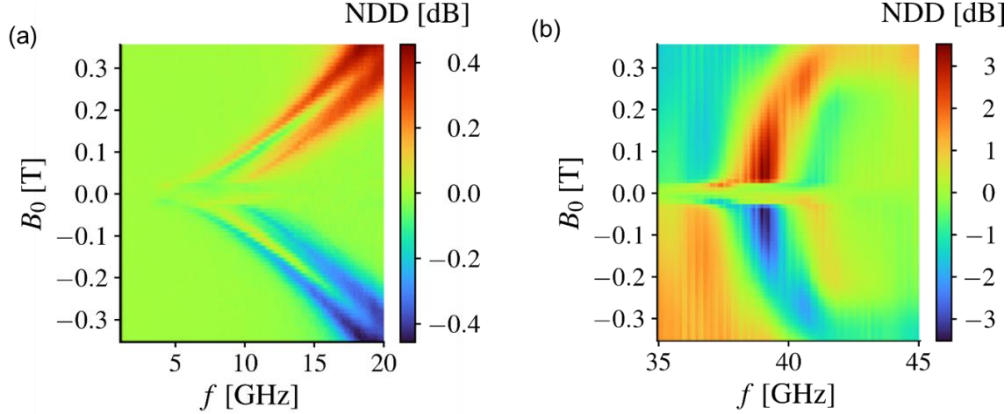

**Figure 4**. **Nonreciprocal directional dichroic signals in microwave transmission spectra.** Around (**a**) low- (3-20 GHz, $f_2$) or (**b**) high- (35-45 GHz, $f_1$) frequency range in various magnetic fields at room temperature with a coplanar waveguide setup (see Methods).

As the slow non-oscillating response in tr-RXD intensities of the (004) reflection (O-3) qualitatively differs for lower and higher laser fluences, we varied the laser fluence on (003) and (004) reflections for a given time delay of 200 ps (see Fig. 5). On the one hand, the intensity of the (004) reflection decreases for increasing fluence up to a threshold laser fluence of approximately 0.7 mJ/cm$^2$ at room temperature, above which the intensity starts to increase and saturates at ~3.1 mJ/cm$^2$ far above its initial value. In contrast, the (003) intensity decreases monotonically over the whole range. On the other hand, the tr-RXD intensities of both the reflections monotonically reduce for increasing laser fluence at 84 K (see blue data points in Fig. 5).

This behavior can be explained by the absorbed energy of the optical excitation. When excited at room temperature, the temperature increases above $T_{MF}$ after equilibration. Above $T_{MF}$, the AFM component is absent, but the FM component is enhanced. Note that the in-plane $M$ in the multiferroic phase does not strongly depend on temperature [22], indicating that any thermal effect on the FM component becomes relevant only for temperatures above $T_{MF}$. This view is supported by the estimated lattice expansion shown in Fig. S1a, consistent with a temperature of ~410 K [32], which is precisely $T_{MF}$. Note also that there is a depth profile of the deposited energy (effective temperature), and the estimated temperature is the mean value within the probe depth. Thus, the top layers have an effective temperature well beyond $T_{MF}$. This scenario explains both the suppression of the (003) reflection and the increase of the (004) reflection.

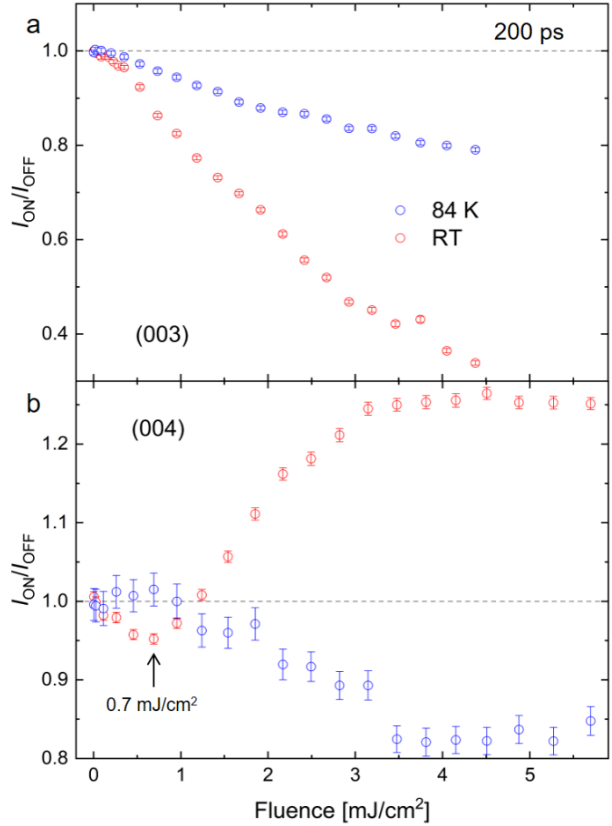

**Figure 5**. **Fluence dependence of (003) AFM and (004) FM reflections.** (**a**) (003) AFM and (**b**) (004) FM measured, taken at room temperature (red) and 84 K (blue), and at 200 ps delay time. Dotted lines indicate normalization baselines.

Understanding the fast dynamics occurring within the first ~200 fs after excitation, as shown in Figs. 2a and 2b, is not straightforward as it is close to the time resolution of the beamline [33]. Nevertheless, tr-RXD intensities for both AFM and FM components show a rapid decrease of magnetic signals within ~200 fs. Such a quick response cannot directly relate to a lattice expansion but is possibly due to an electronic change [34]. To gain more information, we collected time-resolved x-ray absorption spectra (tr-XAS) in fluorescence mode. Figure 6c shows tr-XAS signals for π polarization below or above the prominent main peak of the Fe $L_3$ edge 0.5 ps after the excitation (see Fig. 6a). At both photon energies, transient signals show a rapid variation within ~200 fs, the same time scale as found in the response of the tr-RXD intensities.

As the fast response with a tiny amplitude is also present in the time-resolved circular dichroic signals of the (004) reflection sensitive to $M$ (Fig. 3b), the quick response is expected to be related to the ultrafast demagnetization. Note that the rapid change in fluorescence signals results in an intermediate state with a long lifetime with a monotonic amplitude increase for increasing laser fluence, as shown in Fig. 6b. This long-lived change in XAS remains up to 500 ps (see Fig. 6d), consistent with a magnetic origin, whereas we expect that an electronic change would be very short-lived [35]. It is likely caused by variations in x-ray magnetic linear dichroism due to demagnetization (from the AFM and/or the FM component), though here the changes present only the variation of the single polarization channel (π). The absence of a clear oscillation with the frequency of $f_2$ implies that this mode is unlikely the simple Kittel mode. However, a more detailed investigation with improved statistics would be required to clarify what exactly causes the tr-XAS signals.

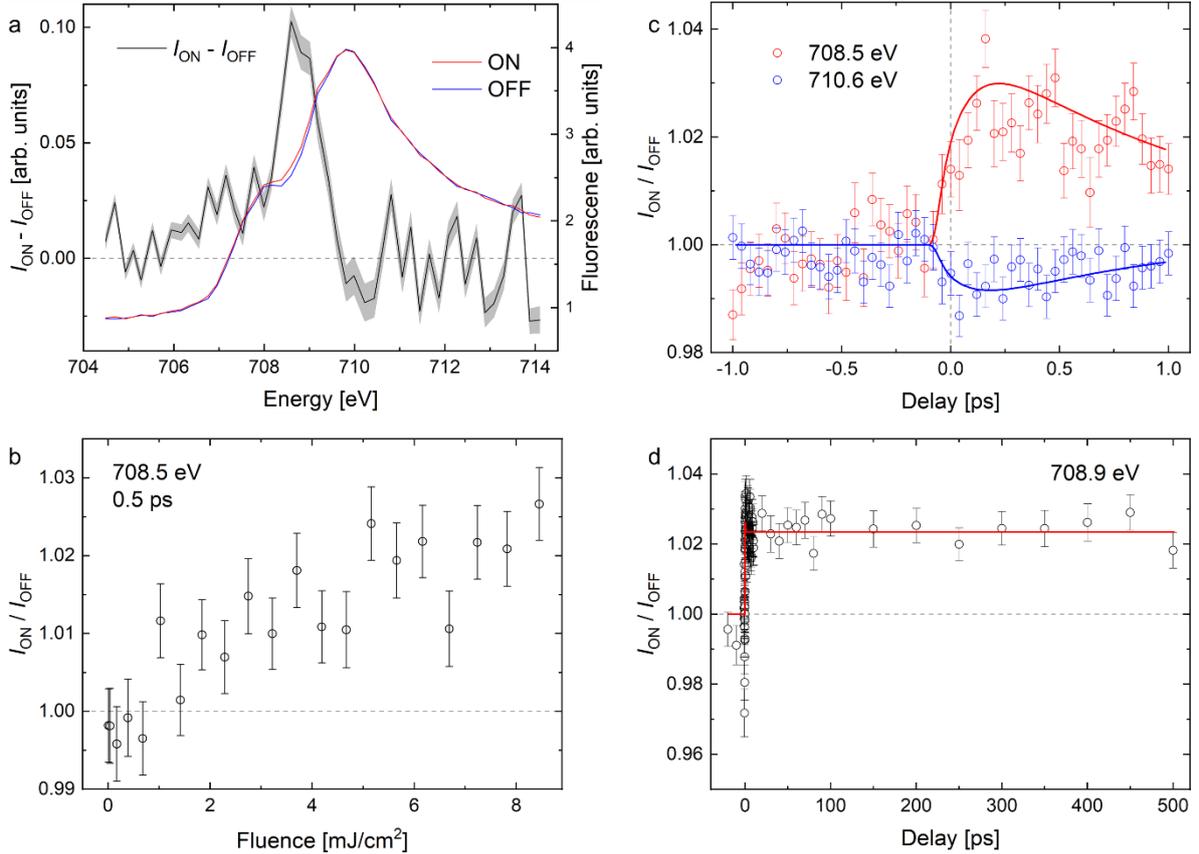

**Figure 6. Time-resolved fluorescence signals.** (**a**) Fluorescence spectra taken with normal incidence of π-polarized x-ray beams for the excited (red, ON, 8.2 mJ/cm$^2$, 0.5 ps after the excitation) and unperturbed (blue, OFF) sample and the difference between them (black). (**b**) Fluence dependence of normalized fluorescence signals taken with the photon energy of 708.5 eV at a time delay of 0.5 ps. Time traces of normalized fluorescence signals (**c**) with a time range around $t_0$ (−1 ps to 1 ps) and (**d**) up to 500 ps. Red, blue, and black symbols were taken with the photon energy of 708.5 eV, 710.6 eV, and 708.9 eV, respectively. The solid curves in (**c**) and (**d**) are represents the best fit to Eq. (1) excluding the oscillation term. Note that the long-lived change indicates an additional recovery term taking more than 500 ps. Horizontal and vertical dotted lines indicate a normalization baseline and $t_0$, respectively.

## DISCUSSION

We have found that there are; (O-1) a rapid response within ~200 fs, observed in tr-RXD intensities of both the (003) and (004) reflections, a time-resolved circular dichroic contrast in (004), as well as in the tr-XAS, (O-2) oscillations with the frequency of $f_1$ ($\approx$ 42.0±1.0 GHz at room temperature) in tr-RXD intensities of both the reflections that are of magnetic origin as supported by the microwave spectra, (O-3) a slow response subsequently following the rapid one, which we interpret as energy transfer from the electronic to the lattice system, (O-4) a recovery that takes longer than 500 ps, and (O-5) oscillations with the frequency of $f_2$ ($\approx$5.7±0.1 GHz) in the time-resolved circular dichroic signals of the (004) reflection most likely representing the lowest-energy magnon mode. Here we discuss (i) the oscillations with $f_1$ representing a coherent magnon mode and (ii) how an above-bandgap excitation launches the magnon mode.

**I. Assignment of the entangled magnon mode**

We assume a simple phenomenological model to describe the experimentally observed entangled magnon mode, i.e., the frequency of $f_1$ with opposite phase oscillations for the two reflections. We consider two order parameters with temporal modulations; an AFM order parameter $S_{\text{AFM}}(t) = S_{\text{AFM}}[1 + \delta e^{i(\omega t - \Phi)}]$ and an FM order parameter $S_{\text{FM}}(t) = S_{\text{FM}}[1 - \delta e^{i(\omega t - \varphi)}]$. Here $S_{\text{AFM}}$ and $S_{\text{FM}}$ are static order parameters of the AFM component and the FM component, respectively, $\delta$ is the amplitude of temporal modulation, $\omega$ is an angular frequency, and $\Phi$ is an arbitrary phase. The (004) reflection at resonance involves scattering processes of magnetic, charge, or orbital (or electric quadrupole) origin. We refer to a parameter describing the latter two scatterings as $S_{\text{CO}}$.

The form factor $\hat{F}$ including the polarization dependence at resonance can be written as a tensor of the form

$$\hat{F} = \begin{pmatrix} F_{\sigma-\sigma'} & F_{\pi-\sigma'} \\ F_{\sigma-\pi'} & F_{\pi-\pi'} \end{pmatrix}. \qquad (2)$$

As the magnetic field in the tr-RXD experiments lies in both the scattering and the basal planes, the magnetic part of the form factor $\mathbf{F}_m$ at (004), i.e., the FM component, is in the scattering plane [$\mathbf{F}_m^{\mathbf{Q}} = \sum_j \mathbf{m}_j \exp(i\mathbf{Q} \cdot \mathbf{r}_j)$, where $\mathbf{m}_j$ and $\mathbf{r}_j$ are the magnetic moment and positional vector of a magnetic atom $j$, respectively, and $\mathbf{Q}$ is the scattering vector.], whereas $\mathbf{F}_m$ at (003), i.e., the AFM component, is normal to $\mathbf{M}$. For π polarized incoming x-rays, magnetic scattering occurs in both the π-σ' and π-π' channels for the (003) reflection. As there is no charge/orbital scattering contribution at (003), the (003) intensities are proportional to $\left|\mathbf{F}_m^{(003)}\right|^2$ and thus $|S_{\text{AFM}}|^2$ as

$$I_{(003)} \approx |S_{\text{AFM}}|^2. \qquad (3)$$

On the contrary, the charge/orbital scattering at (004) appears in the π-π' channel for π polarized incoming x-rays (see Supplementary Information), whereas magnetic scattering appears in the π-σ' channel because $\mathbf{F}_m$ is in the scattering plane. Due to the orthogonal relation between the π-π'channel (charge/orbital scattering) and π-σ' channel (the magnetic scattering), the (004) intensities are

$$I_{(004)} \approx |S_{FM}|^2 + |S_{CO}|^2 = I_{FM} + I_{CO}. \qquad (4)$$

Tr-RXD intensities of the reflections are

$$I_{(003)}(t) \approx |S_{AFM}(t)|^2 = S^*_{AFM}S_{AFM}[1 + \delta e^{-i(\omega t - \varphi)}][1 + \delta e^{i(\omega t - \varphi)}]$$
$$\approx I_{(003)}[1 + 2\delta \cos(\omega t - \varphi)] \qquad (5)$$

and

$$I_{(004)}(t) \approx |S_{FM}(t)|^2 + |S_{CO}|^2 = S^*_{FM}S_{FM}[1 - \delta e^{-i(\omega t - \varphi)}][1 - \delta e^{i(\omega t - \varphi)}] + |S_{CO}|^2$$
$$\approx I_{FM}[1 - 2\delta \cos(\omega t - \varphi)] + I_{CO} = I_{(004)} - 2\delta I_{FM} \cos(\omega t - \varphi). \qquad (6)$$

Equations (5) and (6) describe an entangled mode of the two sublattices and reproduce the experimental observation that is of a single frequency but opposite phase between the (003) reflection and (004) reflection. The initial phase of the oscillations shown in Fig. 2 results in a positive sign of $\delta$, representing an increase of the AFM component subsequent to the excitation. Although changes in magnetic exchange interactions through lattice thermalization can launch magnetic excitations, it should result in a negative sign of $\delta$ due to the reduction of the AFM component at higher temperatures.

Using an effective magnetic Hamiltonian and solving the LLG equation for the hexaferrite, Chun and coworkers obtained two $k = 0$ magnon modes with different frequencies [26]. They found that the two magnetic moments in adjacent L blocks or S blocks have antiphase motions in the magnon modes. The parameters that quantitatively reproduce the higher-frequency electromagnon mode observed by THz-TDS at low temperature (~1 THz at 20 K), provide a slower electromagnon mode frequency of ~58 GHz. This frequency is close to $f_1$ (≈46.3±0.3 GHz at 84 K) found in the tr-RXD intensities. This mode softens when increasing temperatures, for example, ~42.0±1.0 GHz at room temperature, resembling the softening of the higher-frequency electromagnon mode by increasing temperatures [26]. This indicates that $f_1$ obtained from the tr-RXD intensities represents the coherent excitation of the slower electromagnon mode with two entangled sublattices as sketched in Fig. 1b that is dominated by the **μ**$_L$ motion. The antiphase motion of magnetic moments in adjacent L blocks or S blocks creates transient $P$ along [001] through exchange striction, similar to the reported electromagnon mode [25,26]. Inducing transient $P$ through exchange striction implies a more intense resonator strength for the electromagnon mode than for the toroidal magnon mode, which modulates static $P$ originated from spin-orbit couplings, consistent with the observed strengths in the NDD signals (see Fig. 4).

**II. Mechanism of magnon excitation**

The known mechanisms for optical magnon excitations are based on a change in effective magnetic-field direction due to; (i) a photomagnetic effect changing magnetic crystalline anisotropy or known as a coherent displacive excitation [11] and (ii) the inverse

Cotton-Mouton effect or known as an impulsive stimulated Raman scattering [18]. The antiphase spin motions require the opposite direction of transient effective magnetic field between adjacent two L blocks. Among the possible tensor components that can change the effective magnetic-field direction in such a way, however, the symmetry of the hexaferrite does not support such a transient change through these photomagnetic effects (see Supplementary Information for details). Furthermore, photothermalization changing magnetic exchange interactions is ruled out, as mentioned above.

Following the intensive laser excitation, ultrafast demagnetization of a FM sublattice occurs as observed in a rapid drop of the transient (004) circular dichroic signal. It has been shown that most of the lost angular momentum can be transferred within 200 fs into the angular momentum of lattice, creating a transverse force that triggers a strain wave through the ultrafast Einstein-de Haas effect [16]. In ferromagnets with uniform $M$, the force only appears at the surface because a microscopic torque for spatially uniform $M$ cancels in bulk. On the other hand, in ferrimagnets, possessing two blocks with opposite $M$ as drawn in Fig. 7a, the force has the same direction at the interface and is therefore maximal there. This will initiate shear waves (see Fig. 7b and Supplementary Information).

There is the magnetic frustration at the interfaces of the two magnetic blocks, L and S, that creates the non-collinear magnetic structure in the hexaferrite family. Chemical substitution and thus angle changes of specific iron-oxygen bonds (see Fig. 1a) alter the magnetic exchange interaction that strongly affects the frustration and correspondingly the magnetic structure. As the launched shear wave is on a timescale much faster (< 200 fs) than the period of the magnetic excitation, it can launch the electromagnon excitation owing to the sensitivity of the magnetic structure due to the tiny displacements of interfacial bonds.

We can estimate the amplitude of the spin precession from the oscillation amplitude of the tr-RXD intensities of the (003) AFM reflection. The fitting results in Fig. 2 yield the oscillation amplitude in normalized intensities as ~3% at room temperature with 0.5 mJ/cm$^2$, and ~5% at 84 K and 1.2 mJ/cm$^2$. This roughly corresponds to a precession angle of $\mu_L$ as ~1.1° and ~1.8° at room temperature and 84 K, respectively (see Supplementary Information for details). These angles are smaller than that in the Y-type hexaferrite $Ba_{0.5}Sr_{1.5}Zn_2(Fe_{1-x}Al_x)_{12}O_{22}$ ($x = 0.08$) manifesting a different excitation mechanism, namely a coherent displacive excitation type, which is allowed by symmetry [10]. However, the amplitudes in angle displacement found in our experiment for the Z-type hexaferrite are comparable to the reported large-amplitude spin dynamics upon electromagnon resonance excitation in TbMnO$_3$ [8].

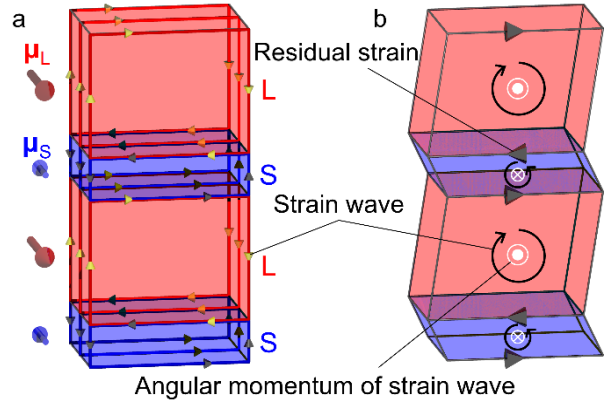

**Figure 7**. **Sketches of the temporal residual strain at the interface of adjacent magnetic blocks due to angular momentum transfer from ultrafast demagnetization of the ferrimagnetic blocks (the ultrafast Einstein-de Haas effect).** Ultrafast demagnetization of the ferrimagnetic component, red and blue arrows in (**a**), resulting in angular momentum transfer, yellow and black arrows in (**a**) and black circle arrows in (**b**), resulting in uncompensated residual strain at the interface of two magnetic blocks as denoted by thick black arrows.

In summary, we investigated the ultrafast magnetization dynamics of the room-temperature multiferroic Z-type hexaferrite $Sr_3Co_2Fe_{24}O_{41}$ that exhibits two magnetic sublattices (ferrimagnetic one and pure antiferromagnetic one), utilizing time-resolved resonant x-ray diffraction upon a femtosecond above-bandgap excitation. Our observations show a coherent entangled magnetic excitation of the two sublattices instantaneously following the optical excitation, which reflects a simultaneous optical control of two characters in a multiferroic material, magnetization and electric polarization, i.e., an electromagnon excitation. We propose that a direct tuning of magnetic frustration due to shear waves launched by ultrafast demagnetization through the Einstein-de Haas effect could be the origin of the coherent excitation of the electromagnon. This novel *non-thermal* mechanism can trigger transient spin modulation in materials with magnetic frustration. The optical manipulation of lattice, based on ultrafast demagnetization and shear waves via the Einstein-de Haas effect, could be, in general, a more competing channel to explore correlations among electronic degrees of freedom and the lattice on an ultrafast timescale.

## Methods

**Time-resolved resonant x-ray diffraction.** Tr-RXD experiments were performed at the RSXS end-station of the SSS beamline in the PAL-XFEL [33]. The photon energies of monochromatic x-ray beams were in the vicinity of the Fe $L_3$ edge (≈ 710 eV). The polarization of the beams was either linear (π) or circular (C+/C−) controlled by the insertion of a magnetized Fe-foil with in-plane *M* (±) at an angle of approximately 45° with respect to the x-ray beam, acting as a circular polarizer [36]. The circular polarization degree estimated from the transmitted intensities is ~80%. Switching **M** of the Fe-foil with an electromagnet reverses the circularly polarized component. The optical pump beams (400 nm) were *p*-polarization and collinear to the probe beam. A single-crystal of $Sr_3Co_2Fe_{24}O_{41}$ was mounted together with a pair of permanent magnets ($\approx$ 0.1 T along [100] and parallel to the scattering plane) on the diffractometer implemented in the end-station (see Fig. 1d). Two avalanche photodiodes collected simultaneous diffraction and fluorescence signals. A gas-monitor detector collected incident x-ray pulse intensities and normalized the signals on a shot-to-shot basis. The penetration depths of the beams are estimated to be ~34 nm and ~13 nm for the pump (obtained from ellipsometry measurements) and for the probe (obtained from x-ray absorption spectroscopy measurements [37]), respectively. The measurements were performed at either room temperature or base temperature of the $LN_2$-cooled cryostat ~84 K. The repetition rates of x-ray probe and optical pump pulses were 60 Hz and 30 Hz, respectively. The length of the x-ray pulses is ~80 fs and is used for convolution of each time trace with the corresponding Gaussian function.

**Microwave transmission spectroscopy.** Room temperature microwave transmission experiments were performed at the Paul Scherrer Institute and employed a coplanar waveguide setup. The same crystal as used for the x-ray experiments was placed on top of the coplanar waveguide, with the [001] direction along the microwave propagation direction. Before the experiments, the sample was processed with magnetic/electric fields, so-called magnetoelectric poling procedures, to create a single-domain multiferroic state as exemplified in Ref. [24]. An electromagnet provided a calibrated magnetic field up to 0.35 T along the [100] direction, which realized the Voigt geometry, typical for NDD studies with coplanar waveguides [30,31]. In the frequency range below 20 GHz, a commercial vector network analyzer recorded the transmitted microwave power. To reach frequencies up to 50 GHz, a custom microwave setup was used. The NDD of transmitted microwaves was obtained by taking the ratio of transmission signals with opposite magnetic field polarity. In particular, data recorded for an up-sweep in the magnetic field was divided by data recorded for the corresponding down-sweep. The Supplementary Information contains further technical details on the microwave setup, transmission data, and NDD at elevated temperatures and a different magnetic field orientation.

**Data availability**

Experimental and model data are accessible from the PSI Public Data Repository [39].

**Acknowledgements**

We thank M. Burian for his advice in data analysis. The time-resolved resonant x-ray diffraction experiments were performed at the RSXS end-station of the SSS beamline in the PAL-XFEL under proposal No. 2020-2nd-SSS-011. The static resonant x-ray diffraction experiments were performed at the X11MA beamline in the Swiss Light Source during in-house access and at the BL17SU in the SPring-8 under proposal No. 20180021. H.U. was supported by the National Centers of Competence in Research in Molecular Ultrafast Science and Technology (NCCR MUST-No. 51NF40-183615) from the Swiss National Science Foundation and from the European Union's Horizon 2020 research and innovation programme under the Marie Skłodowska-Curie Grant Agreement No. 801459 – FP-RESOMUS. H.J. acknowledges the support by the National Research Foundation grant funded by the Korea government (MSIT) (Grant No. 2019R1F1A1060295). S.H.C. was supported by National Research Foundation of Korea (2019R1C1C1010034 and 2019K1A3A7A09033399). N. O. H. was supported by the Swiss National Science Foundation (No. 200021_169017). A.D. acknowledges funding from the PSI-internal CROSS initiative. T.K. and Y.T. were supported by JSPS KAKENHI Grant Number JP19H00661.

**Author contributions**

H.U. and U.S. conceived and designed the project. H.U. and T.K. prepared a single crystal of $Sr_3Co_2Fe_{24}O_{41}$. H.U., N.O.H., Y.T., and U.S. performed synchrotron measurements to characterize the sample. S.F. fabricated the circular polarizer. H.U., H.J., S.H.C., H.-D.K., and U.S. performed the time-resolved resonant x-ray diffraction experiment, and H.U. analyzed the experimental data. S.-Y.P. established the data acquisition system for the beamline experiment. M.K. prepared the optical laser pump setup. V.O. and M.S. measured optical ellipsometry of the sample. A.D. built up the microwave spectroscopy setup and performed the experiment. H.U., A.D., and U.S. interpreted the experimental results and wrote the manuscript. All authors contributed to its final version.

Supplementary Information for

# Optical excitation of electromagnons in hexaferrite

Hiroki Ueda[1,*], Hoyoung Jang[2], Sae Hwan Chun[2], Hyeong-Do Kim[2], Minseok Kim[2], Sang-Youn Park[2], Simone Finizio[1], Nazaret Ortiz Hernandez[1], Vladimir Ovuka[3], Matteo Savoini[3], Tsuyoshi Kimura[4], Yoshikazu Tanaka[5], Andrin Doll[1], and Urs Staub[1,*]

[1]*Swiss Light Source, Paul Scherrer Institute, 5232 Villigen-PSI, Switzerland.*

[2] *PAL-XFEL, Pohang Accelerator Laboratory, Pohang, Gyeongbuk 37673, South Korea.*

[3] *Institute for Quantum Electronics, Physics Department, ETH Zurich, 8093 Zurich, Switzerland.*

[4] *Department of Advanced Materials Science, University of Tokyo, Kashiwa, Chiba 277-8561, Japan.*

[5] *RIKEN SPring-8 Center, Sayo, Hyogo 679-5148, Japan.*

[*] To whom correspondence should be addressed: hiroki.ueda@psi.ch and urs.staub@psi.ch

## Contributions of orbital scattering

In general, scattering from an aspherical electron density created by the occupation valence electrons in the orbitals (orbital scattering) can contribute to diffraction intensities at resonance. Here, we perform a symmetry analysis to clarify if orbital scattering contributes to the (003) and (004) reflections. Orbital scattering is described by the anisotropic x-ray susceptibility tensor $\hat{f}$. The tensor elements are restricted by the local (site) symmetry of the resonant atom, and the sum of the respective tensors with their phase factor gives a form factor and intensity [1].

We denote the symmetric tensor $\hat{f}$ as

$$\hat{f} = \begin{pmatrix} f_{xx} & f_{xy} & f_{xz} \\ f_{xy} & f_{yy} & f_{yz} \\ f_{xz} & f_{yz} & f_{zz} \end{pmatrix} \qquad (S1)$$

in a Cartesian coordinate system where $x$ is along [100], $y$ is along [120], and $z$ is along [001]. There are 10 Fe sites in a Z-type hexaferrite [2], and their local symmetries are summarized in Table S1. All Fe sites except the Fe4 and Fe8 sites have the three-fold rotational symmetry $C_3$ along [001]. The local symmetry requires the relation $\hat{f} = C_3 \hat{f} C_3^{-1}$, which results in $f_{xx} = f_{yy}$ and $f_{xy} = f_{yz} = f_{zx} = 0$. The symmetry-adopted $\hat{f}$ is then

$$\hat{f} = \begin{pmatrix} f_{xx} & 0 & 0 \\ 0 & f_{xx} & 0 \\ 0 & 0 & f_{zz} \end{pmatrix}. \qquad (S2)$$

The Fe4 and Fe8 sites contain the mirror symmetry $m$ that is along <100>. The symmetry-adopted tensor follows the relation $\hat{f} = m\hat{f}m^{-1}$ and is represented as

$$\hat{f} = \begin{pmatrix} f_{xx} & 0 & 0 \\ 0 & f_{yy} & f_{yz} \\ 0 & f_{yz} & f_{zz} \end{pmatrix}. \qquad (S3)$$

For the sites lacking local $C_3$ symmetry, i.e., the Fe(4) and Fe(8), there are three atoms at the same $z$ coordinate due to the global $C_3$ symmetry. For (00$L$) reflections, only the $z$ coordinate is relevant, allowing to define an average $\hat{f}$ for these three atoms that is equivalent to Eq. S2.

As the residual $\hat{f}$'s are isotropic in the basal plane, the form factor $\hat{F}$ for the (003) reflection is zero while that for the (004) reflection is

$$\hat{F} = \begin{pmatrix} F_{xx} & 0 & 0 \\ 0 & F_{xx} & 0 \\ 0 & 0 & F_{zz} \end{pmatrix}. \qquad (S4)$$

Equation S4 implies that orbital scattering appears only in the polarization-unrotated channels, i.e., σ-σ' and π-π' and is independent of the azimuthal angle.

Table S1. Ten Fe sites in a Z-type hexaferrite with their Wyckoff positions and local symmetries, referring to Ref. [2].

| Fe site | Wyckoff position | Local symmetry |
|---|---|---|
| Fe(1) | 2a | $\bar{3}m.$ |
| Fe(2) | 4f | 3m. |
| Fe(3) | 4e | 3m. |
| Fe(4) | 12k | .m. |
| Fe(5) | 4e | 3m. |
| Fe(6) | 4f | 3m. |
| Fe(7) | 4f | 3m. |
| Fe(8) | 12k | .m. |
| Fe(9) | 4f | 3m. |
| Fe(10) | 2c | $\bar{6}m2$ |

## Lattice expansion due to photothermalization

To test if the Bragg peak position (2$\theta$) shifts due to heat-driven lattice expansion we collected (00$L$) diffraction profiles in the vicinity of the (004) reflection at several delay times following laser excitation. Figure S1 shows peak shifts extracted from Gaussian fits of the (004) reflection profiles reflecting possible changes in the lattice constant $c$. There is a tiny change around ~4 ps after the laser excitation, and recovery sets in on a timescale of 500 ps subsequently following the expansion. These peak shifts are more than an order of magnitude smaller than the peak width and therefore do not result in observable variation in the intensity of the peak maxima used in the traces. The peak width of the (004) reflection gets broader by the laser excitation again at ~4 ps after the laser excitation (see Fig. S1). However, these variations are clearly too small to affect the observed oscillations in tr-RXD.

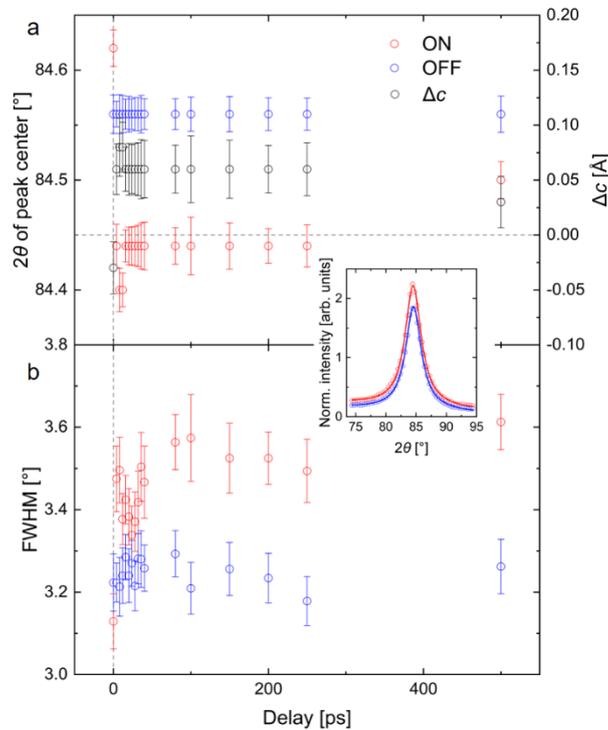

**Figure S1**. **Fitted results of 2$\theta$/$\theta$ profiles around (004) measured**. (**a**) Peak center in 2$\theta$ and (**b**) full width at half maximum (FWHM) as a function of time delay. Red and blue plots indicate data for the excited (ON) and unperturbed (OFF) samples, respectively, at room temperature, whereas black plots indicate a change in the lattice parameter $c$ by the excitation. The inset shows 2$\theta$/$\theta$ profiles at 40 ps of time delay. Laser fluence is 5.7 mJ/cm$^2$, and dotted lines indicate either a baseline or $t_0$.

**Comparison with other known mechanisms of optically excited coherent magnons**

Here we discuss the observed coherent magnon with the other known mechanisms, i.e., (i) a photomagnetic effect changing magnetic crystalline anisotropy or known as a coherent displacive excitation [3] and (ii) the inverse Cotton-Mouton effect or known as an impulsive stimulated Raman scattering [4]. These mechanisms are based on a change in effective magnetic-field direction since for an effective magnetic field $\mathbf{H}_{\mathrm{eff}}$ along the initial magnetization $\mathbf{M}$, the Landau-Lifshitz equation is

$$\frac{d\mathbf{M}}{dt} = -\gamma \mathbf{M} \times \mathbf{H}_{\mathrm{eff}} = \mathbf{0}, \qquad (S5)$$

where $\gamma$ is the gyroscopic ratio. Therefore, we consider if the mechanisms can generate $\mathbf{H}_{\mathrm{eff}}$ that changes the magnetic-field direction. Besides, the magnon mode involves two-order larger motions for $\boldsymbol{\mu}_L$ than $\boldsymbol{\mu}_S$, and so we more specifically consider if the mechanisms can generate $\mathbf{H}_{\mathrm{eff}}$ driving $\boldsymbol{\mu}_L$. To launch the antiphase motion of magnetic moments, $\mathbf{H}_{\mathrm{eff}}$ needs to point in opposite directions between adjacent L blocks.

(i) The photo-induced anisotropy field $\delta\mathbf{H}_i$ through a coherent displacive excitation is phenomenologically described as

$$\delta\mathbf{H}_i(0) = \hat{\chi}_{ijkl}\mathbf{E}_j(\omega)\mathbf{E}_k^*(\omega)\mathbf{M}_l(0), \qquad (S6)$$

where $\hat{\chi}_{ijkl}$ is a fourth-rank polar tensor that is symmetric for $j$ and $k$, $\mathbf{E}$ is the electric field of light [3]. We use the Cartesian coordinate where $x$ // [100] (= 1), $y$ // [120] (= 2), and $z$ // [001] (= 3). In our experimental setup, the scattering plane is spanned by the $z$-$x$ plane, $\boldsymbol{\mu}_L$ is in the $x$-$y$ plane, and the pump beam is $p$-polarized. With this condition, possible components that explain the antiphase spin motion are twelve components among $\hat{\chi}_{ijkl}$ ($\chi_{2111}$, $\chi_{2131}$, $\chi_{2331}$, $\chi_{3111}$, $\chi_{3131}$, $\chi_{3331}$, $\chi_{1112}$, $\chi_{1132}$, $\chi_{1332}$, $\chi_{3112}$, $\chi_{3132}$, and $\chi_{3332}$).

The symmetry of the L block belongs to point group $\bar{6}m2$ and allows only $\chi_{3131}$ among the possible components. However, the symmetry postulates the component invariant between two adjacent L blocks, connected by the two-fold symmetry along [100]. Therefore, the component can give only an in-phase but not an antiphase spin motion between two adjacent L blocks.

(ii) The inverse Cotton-Mouton effect in a magnetic media is due to a photo-induced modulation in dielectric permittivity that is proportional to $\mathbf{M}$ and is phenomenologically described as

$$\mathbf{H}_i^{\mathrm{ICME}}(0) = \hat{g}_{ijkl}\mathbf{E}_j(\omega)\mathbf{E}_k^*(\omega)\mathbf{M}_l(0). \qquad (S7)$$

Here $\hat{g}_{ijkl}$ is a fourth-rank polar tensor that is symmetric for $j$ and $k$ and represents a photo-induced modulation in the permittivity of the media, and $\mathbf{H}_i^{\mathrm{ICME}}$ is the magnetic field induced by the Cotton-Mouton effect [5]. As $\hat{g}_{ijkl}$ has the same intrinsic symmetry with $\hat{\chi}_{ijkl}$, the inverse Cotton-Mouton effect can also not explain the observed antiphase spin motion.

**Estimation of strain at the interface of magnetic blocks**

(i) Local symmetry of magnetic blocks and elastic stiffness tensor

At first, we discuss the local symmetry of the magnetic blocks to find the elastic stiffness tensor components that are allowed by symmetry. The crystal structure of a Z-type hexaferrite comprises two magnetic blocks, an S block and an L block, as shown in Fig. 1(a). An S block locates at $z = 0$ or $1/2$ and is alternating with an L block located at $z = 1/4$ or $3/4$. The space group of the crystal structure is $P6_3/mmc$, and the point group is $6/mmm$. The point group of an S block is $\bar{3}m$, while that of an L block is $\bar{6}m2$. Note that these point groups are subgroups of $6/mmm$. The symmetry-adapted elastic stiffness tensor $C_{ij}$ of the point groups are

$$C_{ij} = \begin{pmatrix} C_{11} & C_{12} & C_{13} & C_{14} & 0 & 0 \\ C_{12} & C_{11} & C_{13} & -C_{14} & 0 & 0 \\ C_{13} & C_{13} & C_{33} & 0 & 0 & 0 \\ C_{14} & -C_{14} & 0 & C_{44} & 0 & 0 \\ 0 & 0 & 0 & 0 & C_{44} & C_{14} \\ 0 & 0 & 0 & 0 & C_{14} & C_{11}/2 - C_{12}/2 \end{pmatrix} \quad (S8)$$

for $\bar{3}m$ and

$$C_{ij} = \begin{pmatrix} C_{11} & C_{12} & C_{13} & 0 & 0 & 0 \\ C_{12} & C_{11} & C_{13} & 0 & 0 & 0 \\ C_{13} & C_{13} & C_{33} & 0 & 0 & 0 \\ 0 & 0 & 0 & C_{44} & 0 & 0 \\ 0 & 0 & 0 & 0 & C_{44} & 0 \\ 0 & 0 & 0 & 0 & 0 & C_{11}/2 - C_{12}/2 \end{pmatrix} \quad (S9)$$

for $\bar{6}m2$ [6].

(ii) Strain through the ultrafast Einstein-de Haas effect

Following Ref. [7], we estimate the strain at the interface of two magnetic blocks. The ultrafast demagnetization causes a transient volume torque density $\boldsymbol{\tau}$ described as

$$\boldsymbol{\tau} = -\frac{1}{\gamma}\frac{d\mathbf{M}}{dt} \quad (S10)$$

[8]. Based on a continuum model, $\boldsymbol{\tau}$ contributes to the off-diagonal components of a magnetization-dependent antisymmetric stress tensor $\widehat{\sigma^M}$ as $\sigma_{12}^M = -\sigma_{21}^M = \tau_3$, $\sigma_{23}^M = -\sigma_{32}^M = \tau_1$, and $\sigma_{31}^M = -\sigma_{13}^M = \tau_2$. The structural dynamics follows the equation of motion with a given stress tensor $\hat{\sigma}$

$$\rho \frac{\partial^2 u_i}{\partial t^2} = \sum_j \frac{\partial \sigma_{ij}}{\partial x_j}, \quad (S11)$$

where $\rho$ is the mass density of the material, $x_j$ ($j = 1, 2, 3$) spans the Cartesian coordinate, and $u_i$ is the displacement along $x_i$. Here $\hat{\sigma}$ contains three contributions written as

$$\sigma_{ij} = \sum_{kl} C_{ijkl}\eta_{kl} + \sigma_{ij}^{M} + \sigma_{ij}^{\text{Ext.}}, \quad (S12)$$

where $\sigma_{ij}^{\text{Ext.}}$ is an external stress tensor component, and $\eta_{kl}$ is the strain described as

$$\eta_{kl} = \frac{1}{2}\left(\frac{\partial u_k}{\partial x_l} + \frac{\partial u_l}{\partial x_k}\right), \quad (S13)$$

where $k, l = 1, 2, 3$.

We describe the tr-RXD setup as $x_1$ // **M** and $x_3$ // [001], namely $\boldsymbol{\tau} = (\tau_1, 0, 0)$. Then only $\sigma_{23}^{M} = -\sigma_{32}^{M} = \tau_1$ are non-zero in $\widehat{\sigma^M}$. Here we ignore the diagonal components of $\widehat{\sigma^M}$ since such components do not contain a torque density. From Eqs. (S10) and (S11), we find, for example,

$$\rho\frac{\partial^2 u_2}{\partial t^2} = \frac{\partial \sigma_{23}}{\partial x_3} = \frac{\partial \tau_1}{\partial x_3} = -\frac{1}{\gamma}\frac{\partial}{\partial x_3}\left(\frac{dM_1}{dt}\right), \quad (S14)$$

indicating that the spatial derivative of the ultrafast demagnetization accelerates the displacement. The displacements are allowed only along $x_2$ // [120] when we assume the pump area is significantly larger than the probe area. To simplify the discussion, we assume that the magnetization and ultrafast demagnetization are both uniform in a magnetic block. Within this assumption, the strain in the interior of a magnetic block is uniform. Besides, the forces from the internal torques cancel everywhere except at the interfaces of the two magnetic blocks, similar to the case of the uniformly magnetized Fe film [7]. The stress through the torque density in a magnetic block appears at both the interfaces with adjacent magnetic blocks, with the same amplitude but opposite sign. Assuming an interface of two magnetic blocks is identical with the surface of the material, we have the same boundary condition at the interfaces with the surface;

$$\sigma_{3j} = 0. \quad (S15)$$

This means at the interfaces

$$\sum_{kl} C_{32kl}\eta_{kl} = -\sigma_{32}^{M} = -\frac{1}{\gamma}\frac{dM_1}{dt}. \quad (S16)$$

The elastic stiffness tensor is $C_{32kl} = C_{4j}$ ($j = k$ if $k = l$, otherwise $j = 9-k-l$). From Eqs. (S9) and (S16), we find only $C_{44} = C_{3232}$ is allowed in the left-hand side of Eq. (S16) for both S and L blocks. Therefore, the transverse strain at the two interfaces that each magnetic block has is

$$\eta_{32} = -\frac{1}{2\gamma C_{3232}}\frac{dM_1}{dt}. \quad (S17)$$

Since the circular dichroic signals on the (004) FM reflection are proportional to magnetization [9], the ultrafast demagnetization is quantitatively estimated from the rapid drop of the signals displayed in Fig. S2. The lost magnetization within 200 fs upon the optical excitation with fluence of 5.3 mJ/cm$^2$ is ~8% of the initial value, corresponding to 0.96 $\mu_B$/f.u. as the magnetization at 1 kG is ~12 $\mu_B$/f.u. (see Fig. S3) We suppose that the timescale and the ratio of angular momentum transfer to the lattice through the ultrafast Einstein-de Haas effect are the same as those of the Fe film, shown in Ref. [7]; 80% of the lost angular momentum transfers to the lattice within 200 fs. There are two interfaces between the magnetic blocks for each formula unit. Thus, each interface acquires 0.38 $\mu_B$/f.u. within 200 fs. Note that our discussion is based on the simple model with uniform magnetization in each magnetic block, meaning a local ferromagnetic structure but not a ferrimagnetic structure as essentially the hexaferrite is.

Using the reported values of $\gamma$ (= $2.76\pi \times 10^6$ /s/T) and $C_{3232}$ (= $3.76 \times 10^{10}$ J/m$^3$) of a similar hexaferrite [10,11], we obtain the strain $\eta_{32} \approx 3.5 \times 10^{-2}$ and the estimated transverse displacements at the interfaces are ~2.2 pm, which is the comparable scale with observed longitudinal phonon modes in, e.g., a Bi or Fe film [12, 7].

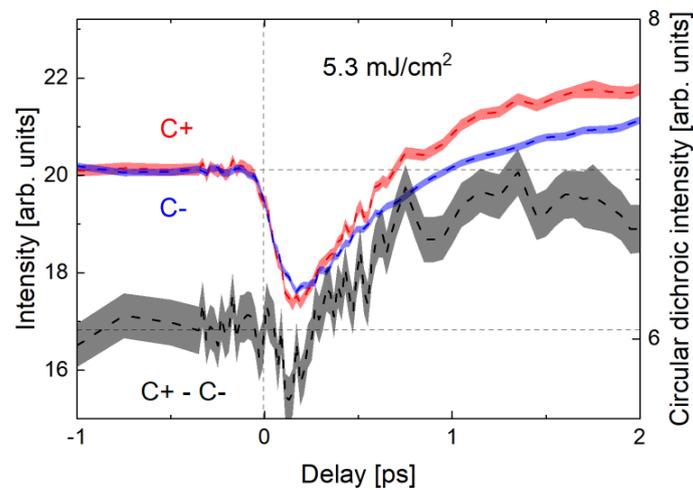

Fig. S2. Time traces of the (004) FM reflection with circular polarization of incident x-rays around $t_0$ measured at room temperature. The fluence is 5.3 mJ/cm$^2$. Red and blue plots are normalized intensities taken with C+ and C− while black plots are the difference between them. Data taken with C− are vertically shifted to match its baseline with data taken with C+.

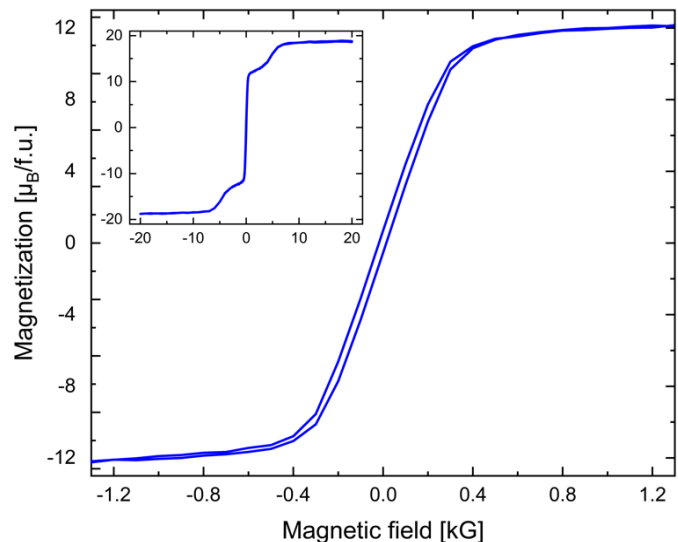

Fig. S3. Magnetization curve measured at room temperature. An applied magnetic field is parallel to the basal plane.

## Estimation of spin-precession angle in the electromagnon

We here quantify the spin-precession angle in the electromagnon mode upon the femtosecond laser excitation. Based on the model shown in Fig. 1(b), the transient magnetic form factor for (003) $\mathbf{F}_m^{(003)}(t)$ is

$$\mathbf{F}_m^{(003)}(t) = 2\begin{pmatrix} 0 \\ -i\mu_L(\cos\Delta\sin\beta + \sin\Delta\cos\beta\cos\omega t) \\ \mu_S\sin\alpha - i\mu_L\sin\Delta\sin\omega t \end{pmatrix}, \quad (S18)$$

where $\mu_{L(S)}$ and $\beta(\alpha)$ are the amplitude of the magnetic moment in an L (S) block and the half opening angle of the conical structure of an L (S) block, respectively (see Fig S4). $\Delta$ stands for the angle between the transient magnetic moment of an L block and its equilibrium direction (see Fig. 1b), and $\omega$ is the angular frequency of the electromagnon mode. Here we use the Cartesian coordinate where $x$ is parallel to $M$, $z$ is parallel to the scattering vector, and $y$ is perpendicular to both $x$ and $z$ (see Fig. 1d for the experimental setup). The wave vectors of incoming x-ray beam (**q**) and outgoing x-ray beam (**q'**) are then

$$\mathbf{q} = \begin{pmatrix} \cos\theta \\ 0 \\ -\sin\theta \end{pmatrix}, \mathbf{q}' = \begin{pmatrix} \cos\theta \\ 0 \\ \sin\theta \end{pmatrix}, \text{ and } \mathbf{q}' \times \mathbf{q} = \begin{pmatrix} 0 \\ \sin 2\theta \\ 0 \end{pmatrix}, \quad (S19)$$

with $\theta$ being the Bragg angle of the (003) reflection. For incoming $\pi$ x-ray polarization, magnetic scattering occurs in the $\pi\sigma'$ ($I_{\pi\sigma'}$) and the $\pi\pi'$ channels ($I_{\pi\pi'}$), which are described by

$$I_{\pi\sigma'} = \left|\mathbf{F}_m^{(003)} \cdot \mathbf{q}\right|^2 \text{ and } I_{\pi\pi'} = \left|\mathbf{F}_m^{(003)} \cdot (\mathbf{q}' \times \mathbf{q})\right|^2, \quad (S20)$$

respectively. Using Eqs. (S18)-(S20), the tr-RXD intensities of (003) are obtained as

$I^{(003)}(t) \approx 4\sin^2\theta(4\mu_L^2\cos^2\theta\sin^2\beta + \mu_S^2\sin^2\alpha) + 2\mu_L^2\sin^2 2\theta \sin 2\Delta \sin 2\beta \cos\omega t,$ (S21)

which is qualitatively the same as Eq. 5. Thus, the normalized intensity $I_{ON}/I_{OFF}$ is

$$\frac{I_{ON}(t)}{I_{OFF}} = 1 + \frac{2\mu_L^2\cos^2\theta \sin 2\Delta \sin 2\beta}{4\mu_L^2\cos^2\theta\sin^2\beta + \mu_S^2\sin^2\alpha}\cos\omega t, \quad (S22)$$

and $\Delta$ is directly defined by the oscillation amplitude extracted from the fits.

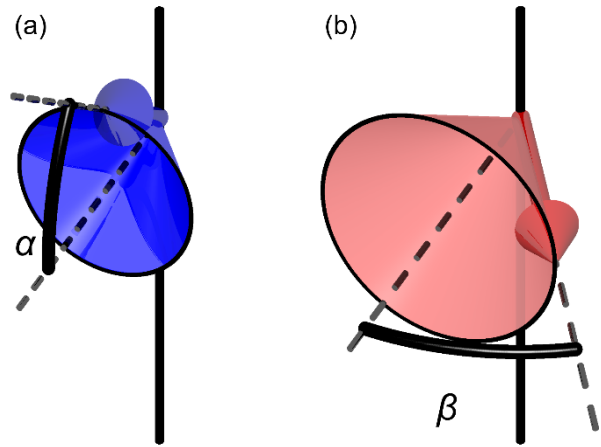

Fig. S4. Definition of the angles in the magnetic structure; (**a**) $\alpha$ [(**b**) $\beta$] is the half opening angle of an S [L] block cone.

## Fitting of tr-RXD intensities

Table S2 shows refined parameters using Eq. 1 convoluted with the time resolution to describe the tr-RXD intensities of the (003) reflection and (004) reflection following the optical excitation with fluence of 0.5 mJ/cm$^2$ at room temperature. The description of each parameter is given in the main text.

Table S2. Refined parameters from the tr-RXD intensities following the optical excitation with fluence of 0.5 mJ/cm$^2$ at room temperature. Strong correlations among parameters prevent obtaining unique parameters for (O-3) and (O-4), which are omitted here.

| Reflection | $A_{rap}$ | $A_{osc}$ | $\tau_{rap}$ [ps] | $\tau_{osc}$ [ps] | $f$ [GHz] | $\varphi$ [°] |
|---|---|---|---|---|---|---|
| (003) | 1.4±0.0×10$^{-2}$ | 3.0±0.0×10$^{-2}$ | 0.25±0.04 | 18±2 | 42.0±1.0 | 179±4 |
| (004) | 8±1×10$^{-2}$ | 4±1×10$^{-2}$ | 0.13±0.02 | 32±6 | 40.8±0.9 | 1±1 |

## Setup for microwave transmission measurements

This section provides details on the experimental setup used for the microwave transmission measurements. As stated in the main text, the sample was mounted on top of a coplanar waveguide (see also Fig. S8 below). For optimum performance up to high microwave frequencies, a commercial coplanar waveguide board was used (Model B4350-30C-50, Southwest Microwave) with suitable 2.92 mm coaxial connectors (Model 1093-01A-5, Southwest Microwave). Experiments in the sub-20 GHz regime utilized a commercial vector network analyzer (Model 8720ES, Agilent Technologies) with output power set to 1 mW. This vector network analyzer (VNA) directly yielded the microwave transmission parameter $S_{21}$ as a function of microwave frequency $f$. In the sub-20 GHz regime, primary experimental data were therefore transmission parameters $S_{21}(f)$ from 1 to 20 GHz recorded at different magnetic field strengths. The magnetic field was stepped in 10 mT steps along a full cycle over its maximum range of ±0.35 T. Since the spectral features in recorded data were quite broad, a smoothing procedure with 240 MHz span was implemented in post-processing to reduce noise contributions.

To reach frequencies up to 50 GHz, a custom setup that implements microwave frequency multiplication was built. In particular, the output frequency of a microwave source (Model EraSynth+, Era Instruments; max. 15 GHz) was multiplied by four. Two frequency doublers (Model TB-973-CY244C +, Mini-Circuits) realized the four-fold frequency multiplication. After passing the first frequency doubler, sufficient drive level and signal purity for the second frequency doubler were provided by a driving amplifier (Model EVAL-HMC383LC4, Analog Devices) followed by a filter (Model TB-883-1832C+, Mini-Circuits). Microwave power levels were measured with a power detector (Model ZV47-K44RMS+, Mini-Circuits). To keep the high-frequency signal path as short as possible, the second frequency doubler and the power detector were directly connected to the coplanar waveguide probe. To ease the voltage readout of the power detector, a modulation scheme with a lock-in amplifier (Model elockin 203, Anaftec Instruments) was implemented: The microwave excitation power was modulated sinusoidally at 910 Hz and the power detector output was demodulated accordingly. The filter time-constant upon demodulation was set to 0.1 s and a settling time of 0.6 s preceded any power level readout. Primary experimental data of this high frequency setup are therefore the transmitted power $P_{tr}$ at a particular frequency and a particular magnetic field. At each microwave frequency, the same magnetic field cycle as for the sub-20 GHz experiments was used.

The high-frequency setup had a lower-frequency cutoff around 30 GHz due to the amplification and filtering stage in-between the two frequency doublers. The upper frequency cutoff was around 51 GHz due to high-frequency limitations of the utilized components. In order to verify the operation of the frequency multiplication chain around 40 GHz, we have temporarily added an extra bandpass filter in front of the coplanar waveguide probe (Model

FB-3700, Marki Microwave). This filter exclusively passed the frequency range from 33 – 41 GHz and approved the features observed in Fig. 4b within the filter passband. In order to access intermediate frequencies between 20 and 30 GHz, the setup was operated with only one of the two frequency doublers.

# Microwave transmission data

In this section, transmission data related to the NDD data in the main text are shown. An important aspect for transmission data is the normalization, which allows compensating for instrumental features. For sub-20 GHz data, normalization was achieved by either data at the highest field for frequencies below 16.7 GHz or data at zero field elsewhere. The particular choice of the frequency crossover at 16.7 GHz was due to the resonance crossing the instrumental baseline. Figure S5a shows the normalized transmission data obtained in this way. Corresponding NDD data is shown in Fig. 4a in the main text. Except for clearly visible fringes in the transmission data, the absorption peaks show a similar field-frequency trend as the NDD data. By comparing the data at positive and negative fields, one can also infer the stronger attenuation for negative fields due to NDD.

To facilitate peak extraction, an additional normalization was applied to the signal attenuation $|S_{21}|_n - 1$: For each frequency $f > 5.75$ GHz, the attenuation was renormalized to the mean peak attenuation for $f > 5.75$ GHz, where peak attenuation refers to the maximum attenuation over the scanned magnetic field cycle. In this way, frequency-dependent fringes were reduced to a constant attenuation value, as is evident in Fig. S5b.

Note that contrary to transmission data, there is no need for any special normalization for NDD extraction, since we extracted NDD by changing the direction of the magnetic field. We have obtained equivalent results by extracting NDD upon changing the microwave propagation direction. However, changing the magnetic field direction yielded superior data quality, since the underlying data originated from exactly the same microwave transmission pathway.

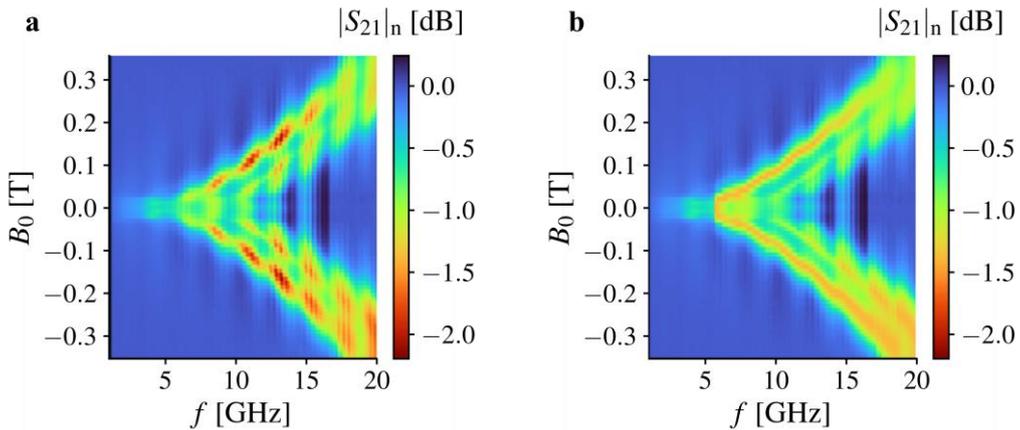

Fig. S5. Normalized transmission data $|S_{21}|_n$ ($f$) at different magnetic fields. (a) Data normalized to instrumental background extracted at highest field ($f < 16.7$ GHz) and zero-field ($f > 16.7$ GHz). (b) Additional scaling of attenuation for fringe suppression in order to ease peak extraction.

In the higher frequency range around 40 GHz, data normalization was less evident since there was no clear absorption peak as in the sub-20 GHz data. Primary un-normalized data are shown in Fig. S6a, where the voltage reading of the power detector was scaled by the nominal detector characteristics, i.e. a 34.4 dB/V slope and an output voltage of 1.0 V for a -5 dBm input signal. For frequencies around 39 GHz, a pronounced asymmetry along the field axis is already visible in this primary dataset. Normalized data are shown in Fig. S6b, where the normalization is the maximum power level over field at each frequency. For negative fields, the heat map resembles ordinary microwave transmission data with a visually traceable field-frequency evolution of the absorption peak. For positive fields, however, the NDD-induced asymmetry renders the absorption peak difficult to trace. In this case, resonant NDD is thus more pronounced than resonant absorption. This is quite different to the sub-20 GHz data, where resonant absorption is predominant and resonant NDD is only visible as a small additional change (Fig. S6a). Note that for the sub-20 GHz data in Fig. S5a, a maximum-based normalization as applied here around 40 GHz yields a qualitatively comparable heat map.

In analogy to the sub-20 GHz data, we have verified that we obtain the same NDD characteristics by reversing the microwave propagation direction. With the setup at 40 GHz, this implied to flip the direction of the coplanar waveguide probe.

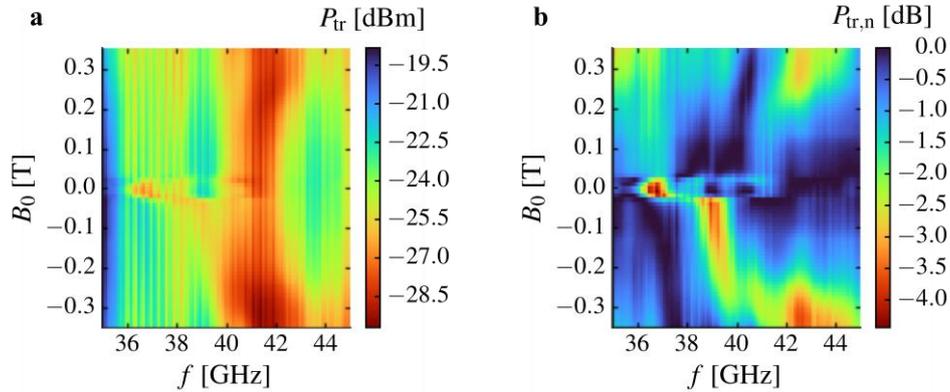

Fig. S6. Transmission power $P_{tr}$ ($f$) around 40 GHz at different magnetic fields. (a) Primary power levels. (b) Data normalized to maximum power level over field at each frequency.

## Microwave data at elevated temperature

In order to support the softening of the 40 GHz mode with temperature observed in time-resolved data, additional microwave transmission spectra at elevated temperature were recorded. For this purpose, a heat gun set to 80 °C was fixed in proximity of the sample. A thermometer brought in contact with the sample displayed a temperature of 76 °C, which we consider as an upper limit for the actual sample temperature.

In the sub-20 GHz regime, softening with temperature was approved by following the evolution of the absorption peak in microwave transmission (Figs. S7a) at room (blue) and elevated temperatures (orange). It is readily seen that for a given magnetic field, heating resulted in mode softening. Moreover, the effect progressed with magnetic field strength. The corresponding NDD spectra at room and elevated temperatures are shown in Figs. S7b and S7c, respectively. For comparison, the evolution in the absorption peak of the related transmission spectra is superimposed onto the NDD spectra as a black curve. Note that sudden jumps in the absorption peak evolution are due to the apparent dual-mode nature of the recorded absorption and NDD spectra. Especially at the largest fields, the branch with slightly higher resonance frequency became dominant (see Fig. S5).

In the 40 GHz regime, temperature effects were traced by following the field-frequency evolution of the most intense NDD (Fig. S7d) at room (blue) and elevated temperatures (orange). Also in this case, the experimental data indicate field-progressive mode softening upon heating, even though less pronounced as in sub-20 GHz data. The corresponding NDD spectra at room and elevated temperature are shown in Figs. S7e and S7f, respectively. When compared to the NDD spectrum in Fig 4b in the main text, NDD was weaker in these experiments. For further reference, the evolution of the NDD peak of the dataset in the main text is shown in green in Fig. S7. We tentatively attribute these differences in NDD to the fact that the sample was re-mounted onto the coplanar waveguide probe in-between these two experiments. We believe that the geometry, orientation and degree of electric polarization of the sample face that is exposed to microwaves are critical parameters for the resultant microwave properties. Importantly, these observations warrant further microwave studies of $Sr_3Co_2Fe_{24}O_{41}$ with a prospect of even larger NDD effects. Likewise, variations in NDD strength were observed among different runs with re-positioned samples in the sub-20 GHz regime. However, NDD strengths stayed within ±0.3 dB with respect to the data showed in Figs. 4 and S7. For sub-20 GHz data acquired at a different magnetic field orientation, however, quite important changes in NDD were observed upon re-positioning (see next section).

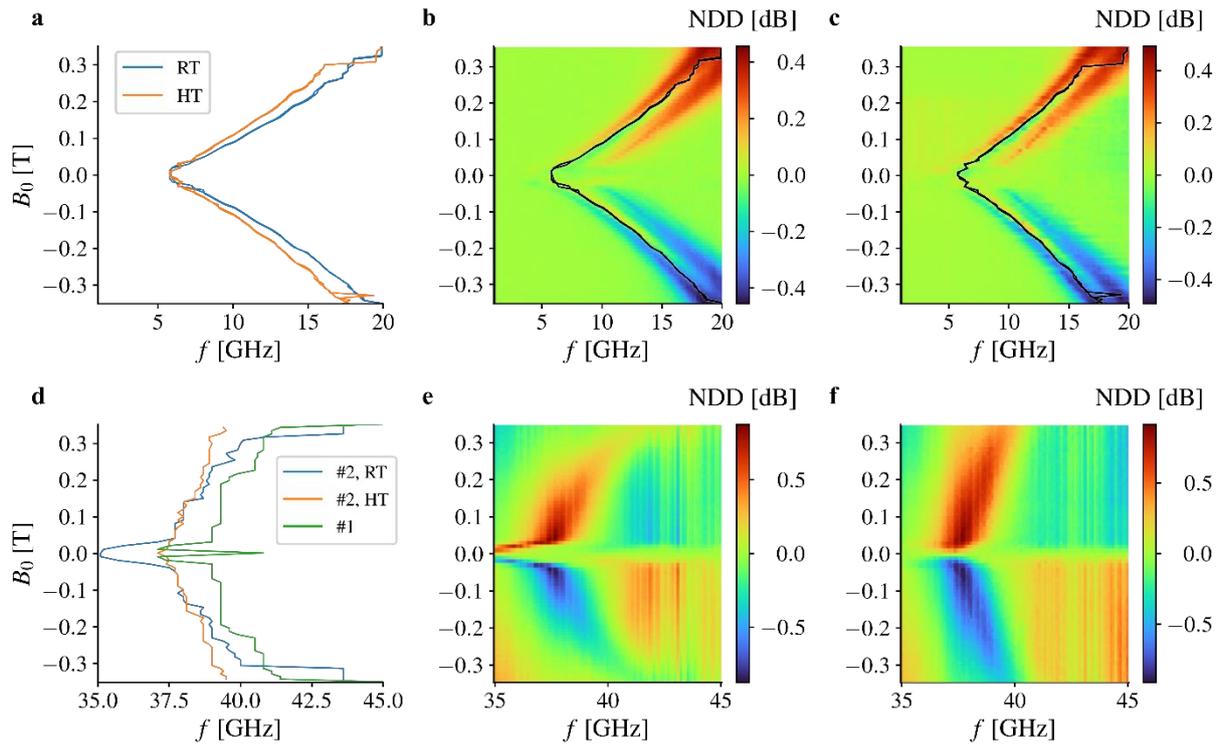

Fig S7. Change in microwave resonances upon heating in the sub-20 GHz (a-c) and 40 GHz (d-f) frequency range. (a) Field-frequency evolution of absorption peak in transmission data at room temperature (blue) and elevated temperature (orange). There are two curves at each temperature, field up- and down-sweep, and peak extraction was according to Fig. S5b. (b,c) NDD data at room and elevated temperatures, respectively. The superimposed black curves are the absorption peaks shown in panel (a). Data in panel (b) is the same as in Fig. 4a in the main text. (d) Field-frequency evolution of resonant NDD peak at room (blue) and elevated temperatures (orange). The green curve depicts the NDD peak for the NDD spectrum shown in Fig. 4b in the main text. (e,f) NDD data at room and elevated temperatures, respectively.

**Microwave data at different field orientation**

In previous sections, ultrafast time-resolved XFEL and frequency-domain microwave transmission experiments were performed with the static magnetic field along the [100] direction. Since the two techniques bear different excitation mechanisms, we also performed microwave experiments with the field along the [120] direction. The orientation of the sample with respect to the field and coplanar waveguide is depicted in Figs. S8a and S8b. The photograph in Fig. S8c shows the actual setup with field along the [100] direction and the microwave propagation direction $k_{\mu w}$ as indicated. Rotation of the magnetic field around [001] was accomplished by fixation of the coplanar waveguide probe at a different angle. Such a rotation does not only change the orientation of the relevant magnetic field in the basal plane, but also changes the relative orientation between static and microwave fields. Figure S8d illustrates schematically the direction of the electric and magnetic microwave fields due to the coplanar waveguide. Changes observed upon sample rotation are thus a combination of (i) anisotropy with respect to the external field within the basal plane and (ii) anisotropy with respect to the microwave excitation fields.

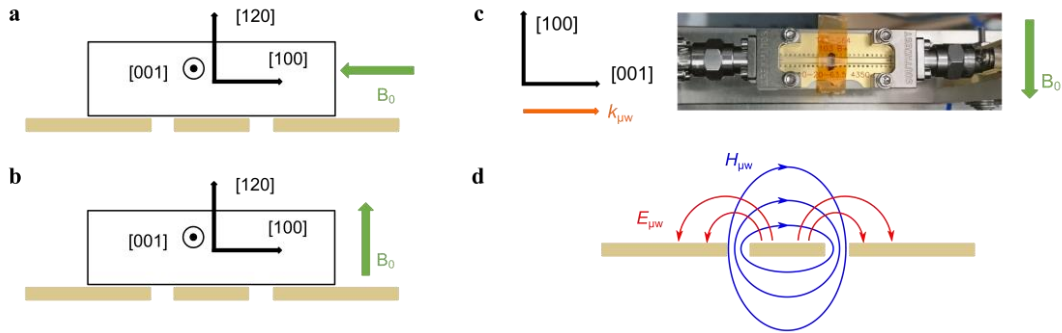

Fig S8. Sample and probe geometry for microwave transmission experiments. (a,b) Schematic layout of sample (black) placed on top of coplanar transmission waveguide cross section (beige) with the external field along the [100] or [120] direction, respectively. (c) Photograph of the sample mounted on top of the coplanar waveguide. The center trace has a width of 43 mils (1.1 mm). (d) Indicative illustration of typical microwave field distribution within the waveguide cross section. The arrows represent field lines of constant intensity.

Figure S9 shows microwave transmission (a,c) and NDD (b,d) in the sub-20 GHz regime with field along the [120] direction. Two different datasets are shown in order to point at the subtle influence of sample positioning and microwave properties. For upper dataset, remarkably large NDD was observed for frequencies around 17.5 GHz. For the lower dataset, the NDD pattern is in qualitative agreement, but at much lower intensity. In accordance to our observations in the 40 GHz regime (Fig. S7), we anticipate the possibility to fine tune the NDD response and reach even larger effects with $Sr_3Co_2Fe_{24}O_{41}$. In contrast to the 40 GHz data, however, sub-20 GHz data feature pronounced resonant microwave absorption that can be compared against each other via the corresponding transmission spectra in panels (a) and (c). Interestingly, these transmission data show quantitatively comparable absorption levels. The apparent differences in the color scale are mainly caused by the different NDD responses that superimpose on normalized transmission data.

As is evident when comparing [Fig. S9](Fig. S9) and [Fig. S5](Fig. S5), there are important changes when flipping the magnetic field direction. To further ease the comparison, the field-frequency evolution of the principal absorption peak from Fig. S5b was superimposed as black line onto transmission spectra in [Fig. S9](Fig. S9). These superimposed lines follow the highest-frequency ridge in the transmission spectra. We thus tentatively conclude that there is a common mode that is excited with the field along either [100] or [120]. With the field along [100], this common mode is the lowest-frequency ridge in the transmission spectrum, while with the field along [120], the common mode becomes the highest-frequency ridge in the transmission spectrum. It is also interesting to note that the with the field along [120], the lowest-frequency ridge has a resonance frequency of 7 GHz at 0.1 T, which is very close to the beating observed in laser-excited time-domain XMCD data in the main text.

For further insight into the origin of the observed anisotropy in microwave transmission, additional experiments with the magnetic field at intermediate orientations between [100] and [120] have been performed (data not shown). These experiments indicated that field orientations along [100] or [120] constitute extremum constellations, from which we tentatively deduce a two-fold rotational symmetry of the apparent anisotropy. As mentioned above, the anisotropy can originate from both anisotropy related to the sample or related to the microwave excitation geometry. Since the sample has hexagonal symmetry in the basal plane, we consider anisotropy with respect to the microwave excitation fields to be the more likely cause for the two-fold rotational symmetry. Within this viewpoint, there is the possibility that laser-excited microwave mode in time-domain experiments with the field along [100] corresponds to the lowest-frequency ridge of the microwave-excited modes with the field along [120]. Given the anisotropy and multitude of microwave-excited modes in the sub-20 GHz regime, however, the microscopic relationship between the microwave- and laser-excited modes is beyond the scope of the current study and subject to further research.

An anisotropy in microwave properties was also observed for the mode around 40 GHz. In this case, the NDD response with field along the [100] direction ([Fig. S10a](Fig. S10a)) changed its characteristic field-frequency evolution when applying the field along the [120] direction ([Fig. S10b](Fig. S10b)). As is readily observed, the resonance was almost field-independent in this configuration. Note that the broader range scans in [Fig. S10](Fig. S10) were performed with a ten times lower resolution in the microwave frequency axis than in other illustrated NDD spectra at 40 GHz.

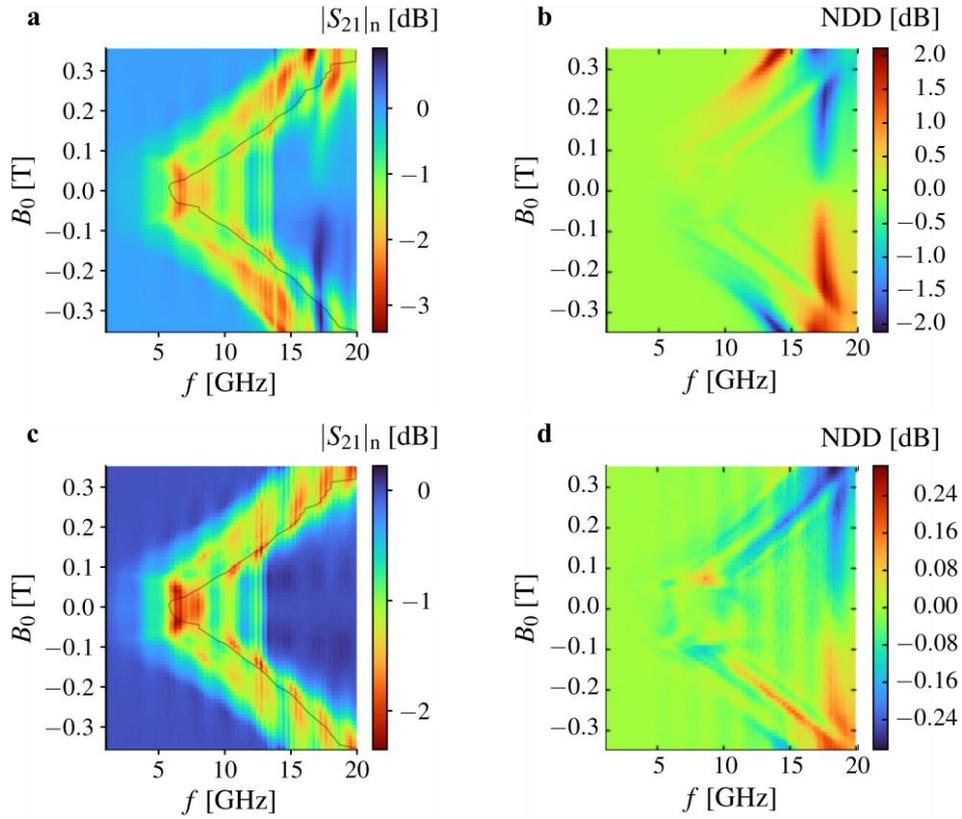

Fig S9. Microwave response in sub-20 GHz regime with magnetic field along [120], showing normalized microwave transmission (a,c) and NDD (b,d) of the same sample mounted twice.

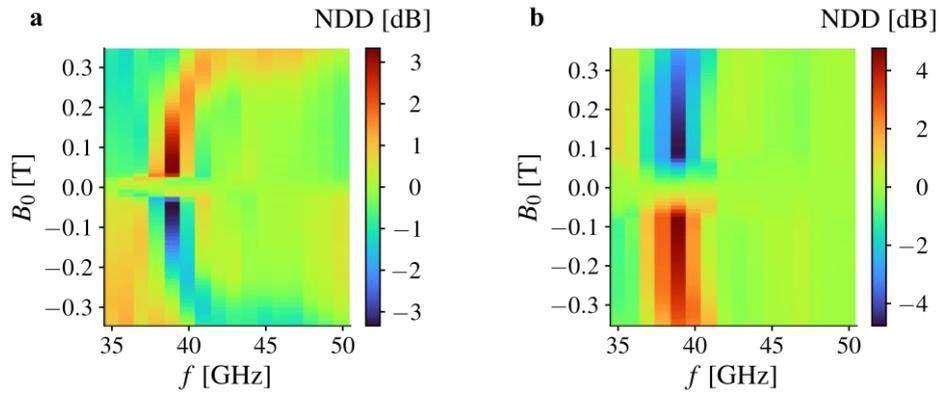

Fig S10. 40 GHz NDD with 1 GHz frequency steps and magnetic field along [100] (a) or [120] (b).